\documentclass[11pt]{article}
\usepackage[margin=1in]{geometry}
\usepackage{float}
\usepackage{flafter}
\usepackage{subfigure}
\usepackage{varwidth}
\usepackage{graphicx}
\usepackage[T1]{fontenc}
\usepackage[justification=centering]{caption}

\begin{document}
\title{A Unified Programming Model for Heterogeneous Computing with CPU and Accelerator Technologies\footnote{The first draft of the first part in this paper was written in June 2018. To the best of the author's knowledge, the author is the first to propose the idea of the unified programming model for Heterogeneous Computing with CPU and Accelerator Technologies.}}
\author{Yuqing Xiong\\
(Computer Science Department, Shanghai Institute of Technology, Shanghai, China)\\
\texttt{yqxiong@sit.edu.cn}}
\date{}

\maketitle

\begin{abstract}
This paper consists of three parts. The first part provides a unified programming model for heterogeneous computing with CPU and accelerator (like GPU, FPGA, Google TPU, Atos QPU, and more) technologies. To some extent, this new programming model makes programming across CPUs and accelerators turn into usual programming tasks with common programming languages, relieves complexity of programming across CPUs and accelerators, and causes programs modularity higher.  It can be achieved by extending file managements in common programming languages, such as C/C++, Fortran, Python, MPI, etc., to cover accelerators as I/O devices. In the second part, we show that all  types of computer systems can be reduced to the simplest type of computer system, a single-core CPU computer system with I/O devices, by the unified programming model. Thereby, the unified programming model can truly build the programming of various computer systems on one API (i.e. file managements of common programming languages), and can make programming for various computer systems  easier. In the third part, we present a new approach to coupled applications computing (like multidisciplinary simulations) by the unified programming model. The unified programming model makes coupled applications computing more natural and simpler since it only relies on its own power to couple multiple applications through MPI. 
\end{abstract}


\section{Introduction}
Heterogeneous computing with CPU and accelerator technologies is widely concerned. However, there are some challenges in the heterogeneous computing. To remove the difficulties, for example, an architecture for unified deep learning with CPU, GPU, and FPGA technologies is presented[1]. The examples of underlying hardware approaches in the architecture are shown in Figure 1 and Figure 2 (extracted from [1]). This is a typical heterogeneous computing architecture with CPU and accelerator technologies.\\

\begin{figure}[htbp]
\begin{minipage}[t]{0.5\linewidth}
\centering
\includegraphics[scale=.65]{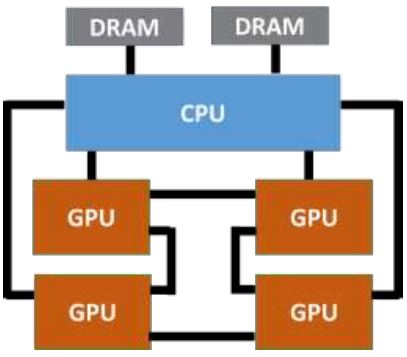}
\caption{GPU Scale out using Vega20 and Epyc\newline (extracted from [1])}
\end{minipage}%
\begin{minipage}[t]{0.5\linewidth}
\centering
\includegraphics[scale=.65]{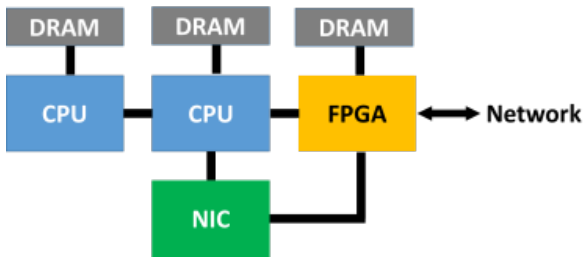}
\caption{FPGA Exploitation\newline (extracted from [1])}
\end{minipage}
\end{figure}

\noindent A key problem for the heterogeneous computing is that a full and seamless programming environment that works across CPUs and accelerators is necessary so that the architecture can work well [1]. However, it seems to difficult to design and build the full and seamless programming environment since it is not easy that communication between application programs of common programming languages for CPUs and programs of programming languages for accelerators are carried out. For example, data movement between MPI for distributed memory parallel programming and CUDA is difficult, it makes programming with MPI+CUDA complicated [2].\\

\noindent To overcome the challenge of communication between CPUs and accelerators in the common programming language world, this paper tries to provide a unified programming model that can work across CPUs, GPUs, FPGAs, and other accelerators (such as Google TPU, Atos QPU, etc.) by extending file managements in common programming languages, such as C/C++, Fortran, Python, MPI, etc., to cover GPUs, FPGAs, and other accelerators (such as Google TPU, Atos QPU, etc.) as I/O devices. To some extent it makes programming across CPUs and accelerators turn into usual programming tasks, and relieve complexity of programming across CPUs, GPUs, FPGAs, and other accelerators since it makes heterogeneous systems turn into homogeneous systems from a certain perspective (accelerators as one kind of I/O devices). Thus the programming model can contribute to improve software productivity for computing across CPUs and accelerators.\\

\noindent To further explain the unified programming model to simplify programming complexity, in this paper, we will show that all  types of computer systems can be reduced to the simplest type of computer system, a single-core CPU computer system with I/O devices, by the unified programming model. Thereby, the unified programming model can truly build the programming of various computer systems on one API (i.e. file managements of common programming languages), and can make programming for various computer systems  easier.\\

\noindent Coupled applications computing (like multidisciplinary simulations) is very important in many fields. It usually needs addtional software to support [3]. In this paper, we will present a new approach to coupled applications computing, which is based on the unified programming model.  The unified programming model makes coupled applications more natural and easier since it only relies on its own power to couple multiple applications through MPI.

\section{A Unified Programming Model for Heterogeneous Computing with CPU and Accelerator Technologies}

\subsection{Accelerators as One Kind of I/O Devices in Common Programming Languages}
File managements in common programming languages are a kind of mechanisms which make application programs access I/O devices easier. An application's access to I/O devices generally involves many tasks carried out by systems, that is, the tasks are accomplished by systems (not by applications) in the name of application access to I/O devices through file managements. This name is the key to making it easy for applications (running on CPUs) to access I/O devices.\\

\noindent Therefore, if we can also regard accelerators as one kind of I/O devices (or one kind of general printers exactly) in common programming languages (although accelerators, such as GPUs, aren't any kind of I/O devices in the usual sense) and extend file managements in common programming languages to cover accelerators as one kind of  I/O devices, the data movement between CPUs and accelerators can also be carried out by systems in the name of application access to I/O devices through the file managements. Thus we can explicitly avoid data movement between CPUs and accelerators in application programs of the common programming languages and make programming across CPUs and accelerators easier. GPUs are taken as an example to illustrate it here.\\

\noindent Let's compare GPUs with HP printers. There are many models (such as LaserJet P1008) for the HP printers, there is a driver for each model, and outputs of printers are to papers. Similarly, ``GPU+CUDA code running on the GPU'' can be regarded as an I/O device (a general printer), its model is the CUDA code, its driver is also the CUDA code, and its output is to CPUs (application processes more exactly).\\

\noindent Let's take an example to illustrate it further, see Figure 3. Assume that P is a program written in a common programming language and runs on a CPU, Q is a program written in CUDA and runs on a GPU. ``The GPU+Q'' is regarded as a general printer, the code of Q is its model, and Q is its deriver. Now P invokes Q, so P sends data to Q (i.e. P sends input to the general printer to print), and Q receives the data, processes the data, and then sends the computational result to P (i.e. the general printer prints the results to P). Thus communication between P and Q is carried out, and P completes the invoking to Q. Figure 3 shows the communication between P and Q.\\

\begin{figure}[htbp]
\centering
\includegraphics[scale=.80]{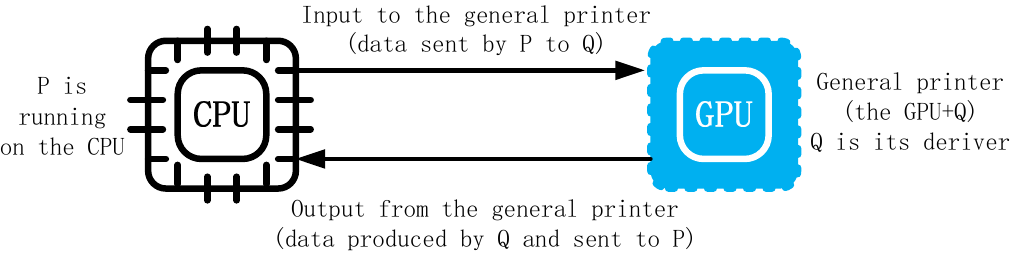}
\caption{Communication between P running on a CPU and Q running on a GPU.}
\end{figure}

\noindent Assume that \textsf{print()} is a function in the extended file management of the common programming language P is written in, D$_{input}$ is the data sent by P to Q, and the computational result sent to P by Q is  ``printed'' on D$_{output}$. The above description of P invoking Q can be expressed by the function as follows:

\begin{center}
\textsf{D$_{output}$=print(the GPU+Q, D$_{input}$)}
\end{center}

\noindent This way, P invoking Q through \textsf{print()} triggers the system to carry out the communication between P in the CPU and Q in GPU implicitly, that is, we realize data movement between application programs of common programming languages on CPUs and programs of CUDA on GPUs in the name of printing the data by the general printer.\\

\noindent In operating systems and common programming languages, such as C/C++, Fortran, Python, etc., all I/O devices (including printers) are regarded as files, so it should be natural that ``GPU+CUDA code running on the GPU'' can also be regarded as files in common programming languages.\\

\noindent Naturally, we should extend file managements in common programming languages, such as C/C++, Fortran, Python, etc., to include accelerators as one kind of I/O devices. When the common programming language is MPI, MPI-I/O should be extended to cover accelerators as one kind of I/O devices, and correspondingly the file management in C should be extended to include accelerators as one kind of I/O devices since MPI is implemented in C.\\

\noindent Thus, we simplify programming across CPUs, GPUs, FPGAs, and other accelerators into usual programming tasks with common programming languages to some extent, and make programming in heterogeneous systems with accelerators easier since it makes heterogeneous systems turn into homogeneous systems from a certain perspective (accelerators as one kind of I/O devices). So this can contribute to improve software productivity for computing across CPUs and accelerators.\\

\noindent By this point, we have established a unified progrmming model for heterogeneous computing with CPU and accelerator technologies. For convenience,  we use UPM as an abbreviation for   ``the Unified Programming Model for heterogeneous computing with CPU and accelerator technologies''.



\subsection{For Multi-core CPUs for OpenMP or Pthread in UPM}

In UPM, for multi-core CPUs where programs based on OpenMP or Pthread run, we can imagine that every multi-core CPU consists of one ``single-core CPU'' for programs of common programming languages and one ``multi-core accelerator'' for programs based on OpenMP or Pthread\footnote[7]{After the first part of this paper appeared (i.e., after UPM was established), Mr. James Reinders at Intel thought that a multi-core CPU could be regarded as an accelerator. Inspired by it, the author thought a multi-core CPU should be imagined as a single-core CPU + a multi-core ``accelerator'' in UPM. The author would like to thank Mr. Reinders althougt he didin't mention UPM. The author think that it should make little sense without UPM that a multi-core CPU can be regarded as an accelerator.}. In the same way, we can regard the ``multi-core accelerators + program codes based on OpenMP/Pthread running on the multi-core accelerators'' as I/O devices, and extend file managements in common programming languages to include them as one kind of I/O devices (i.e. files). Thus to some extent we also simplify programming model ``common programming languages + OpenMP/Pthread'' into usual programming with common programming languages.\\

\noindent Thus, UPM allows OpenMP/Pthread-based code to be separated from the entire program, resulting in higher modularity. 

\subsection{For Clusters of CPUs for MPI in UPM}

In UPM, for clusters of CPUs where programs based on MPI run, we imagine that every cluster of CPUs is composed of one CPU for programs of common programming languages and one ``cluster-of-CPU accelerator'' for programs based on MPI. We use similar ideas as in section 2.2 and regard the  ``cluster-of-CPU accelerator + program codes based on MPI running on the cluster-of-CPU accelerator'' as an I/O device, and extend file managements in common programming languages to include it as one kind of I/O devices (i.e. files).  \\

\noindent According to this, UPM may make  coupled applications computing (like multidisciplinary simulations) more natural and easier, see section 4 below.

\section{Computer Systems under the View of UPM}

So far, there are many kinds of computer systems in the world. We roughly divide these computer systems into several types according to whether CPUs are single-core and whether there is a cluster of CPUs and whether there are accelerators. Although some computer systems may not belong to any type, this does not affect the results of our discussion in this paper. Assume that all the computer systems are attached with I/O devices. We give a description of every type of computer system as follows.

\begin{center}
\centering
\begin{tabular}{|l||l|}\hline
\textbf{Type}&\textbf{Description}\\
\hline\hline
\uppercase\expandafter{\romannumeral1}&a single-core CPU\\
\hline
$\rm\uppercase\expandafter{\romannumeral1}^+$&a single-core CPU with accelerators\\
\hline
\uppercase\expandafter{\romannumeral2}&a multi-core CPU\\
\hline
$\rm\uppercase\expandafter{\romannumeral2}^+$&a multi-core CPU with accelerators\\
\hline
\uppercase\expandafter{\romannumeral3}&a cluster of single-core CPUs\\
\hline
$\rm\uppercase\expandafter{\romannumeral3}^+$&a cluster of single-core CPUs with accelerators\\
\hline
\uppercase\expandafter{\romannumeral4}&a cluster of mutil-core CPUs\\
\hline
$\rm\uppercase\expandafter{\romannumeral4}^+$&a cluster of multi-core CPUs with accelerators\\
\hline
\end{tabular}
\end{center}

\noindent For convenience,  we use $\Longrightarrow$ as   ``can be reduced to''. A $\stackrel{UPM}\Longrightarrow$ B represents that A can be reduced to B by UPM.

\subsection{All Types of Computer Systems $\stackrel{UPM}\Longrightarrow$ Type-\uppercase\expandafter{\romannumeral1} Computer Systems}

In this section, we show that every type of computer system can be reduced to a type-\uppercase\expandafter{\romannumeral1} computer system by UMP, i.e., any type of computer system $\stackrel{UPM}\Longrightarrow$ a type-\uppercase\expandafter{\romannumeral1} computer system.

\subsubsection{Type-\uppercase\expandafter{\romannumeral1} Computer Systems}
A type-\uppercase\expandafter{\romannumeral1} computer system contains only a single-core CPU and is the simplest computer system. Its programming model is common programming languages, such as C/C++, Fortran, Python, Java and so on, and is the simplest programming method.

\subsubsection{Type-$\rm\uppercase\expandafter{\romannumeral1}^+$ Computer Systems $\stackrel{UPM}\Longrightarrow$ Type-\uppercase\expandafter{\romannumeral1} Computer Systems}

A type-$\rm\uppercase\expandafter{\romannumeral1}^+$ computer system is composed of type-\uppercase\expandafter{\romannumeral1} computer system and accelerators. According to section 2.1, the accelerators in the type-$\rm\uppercase\expandafter{\romannumeral1}^+$ computer system can be regarded as I/O devices (general printers exactly), and then as files by UMP. Thus, the accelerators in the type-$\rm\uppercase\expandafter{\romannumeral1}^+$ computer system  ``disappear'', and the type-$\rm\uppercase\expandafter{\romannumeral1}^+$ computer system turns into a type-\uppercase\expandafter{\romannumeral1} computer system, i.e., the type-$\rm\uppercase\expandafter{\romannumeral1}^+$ computer system $\stackrel{UPM}\Longrightarrow$ a type-\uppercase\expandafter{\romannumeral1} computer system.

\subsubsection{Type-\uppercase\expandafter{\romannumeral2} Computer Systems $\stackrel{UPM}\Longrightarrow$ Type-\uppercase\expandafter{\romannumeral1} Computer Systems}

In a type-\uppercase\expandafter{\romannumeral2} computer system, the CPU is a multi-core CPU. According to section 2.2, the type-\uppercase\expandafter{\romannumeral2} computer system is composed of a type-\uppercase\expandafter{\romannumeral1} computer system and a ``multi-core accelerator''. The ``multi-core accelerator'' can be regarded as an I/O device (a general printer exactly), and then a file by UPM. Thus, the ``multi-core accelerator'' also ``disappears'', and the type-\uppercase\expandafter{\romannumeral2} computer system turns into a type-\uppercase\expandafter{\romannumeral1} computer system, i.e., the type-\uppercase\expandafter{\romannumeral2} computer system $\stackrel{UPM}\Longrightarrow$ a type-\uppercase\expandafter{\romannumeral1} computer system. 

\subsubsection{Type-$\rm\uppercase\expandafter{\romannumeral2}^+$ Computer Systems $\stackrel{UPM}\Longrightarrow$ Type-\uppercase\expandafter{\romannumeral1} Computer Systems}

A type-$\rm\uppercase\expandafter{\romannumeral2}^+$ computer system is composed of a type-\uppercase\expandafter{\romannumeral2} computer system and accelerators. According to the same reason as in section 3.1.2, the type-$\rm\uppercase\expandafter{\romannumeral2}^+$ computer system $\stackrel{UPM}\Longrightarrow$ a type-\uppercase\expandafter{\romannumeral2} computer system. By section 3.1.3,  the type-\uppercase\expandafter{\romannumeral2} computer system $\stackrel{UPM}\Longrightarrow$ a type-\uppercase\expandafter{\romannumeral1} computer system. Therefore, the type-$\rm\uppercase\expandafter{\romannumeral2}^+$ computer system $\stackrel{UPM}\Longrightarrow$ the type-\uppercase\expandafter{\romannumeral1} computer system.

\subsubsection{Type-\uppercase\expandafter{\romannumeral3} Computer Systems $\stackrel{UPM}\Longrightarrow$ Type-\uppercase\expandafter{\romannumeral1} Computer Systems}

There is a cluster of single-core CPUs in a type-\uppercase\expandafter{\romannumeral3} computer system, and programs based on MPI are for it. According to section 2.3, the type-\uppercase\expandafter{\romannumeral3} computer system is composed of a single-core CPU and  a ``cluster-of-single-core-CPU accelerator''. By UPM, the ``cluster-of-single-core-CPU accelerator'' can be regarded as an I/O device (a general printer exactly), and then a file. Thus, the ``cluster-of-single-core-CPU accelerator'' also ``disappears'', and the type-\uppercase\expandafter{\romannumeral3} computer system turns into a type-\uppercase\expandafter{\romannumeral1} computer system, i.e., the type-\uppercase\expandafter{\romannumeral3} computer system $\stackrel{UPM}\Longrightarrow$ a type-\uppercase\expandafter{\romannumeral1} computer system.

\subsubsection{Type-$\rm\uppercase\expandafter{\romannumeral3}^+$ Computer Systems $\stackrel{UPM}\Longrightarrow$ Type-\uppercase\expandafter{\romannumeral1} Computer Systems}

A type-$\rm\uppercase\expandafter{\romannumeral3}^+$ computer system is composed of a type-\uppercase\expandafter{\romannumeral3} computer system and accelerators. According to the same reason as in section 3.1.2, the type-$\rm\uppercase\expandafter{\romannumeral3}^+$ computer system $\stackrel{UPM}\Longrightarrow$ a type-\uppercase\expandafter{\romannumeral3} computer system. By section 3.1.5, the type-\uppercase\expandafter{\romannumeral3} computer system $\stackrel{UPM}\Longrightarrow$ a type-\uppercase\expandafter{\romannumeral1} computer system. Therefore, the type-$\rm\uppercase\expandafter{\romannumeral3}^+$ computer system $\stackrel{UPM}\Longrightarrow$ the type-\uppercase\expandafter{\romannumeral1} computer system.

\subsubsection{Type-\uppercase\expandafter{\romannumeral4} Computer Systems $\stackrel{UPM}\Longrightarrow$ Type-\uppercase\expandafter{\romannumeral1} Computer Systems}

In a type-\uppercase\expandafter{\romannumeral4} computer system, the CPUs are multi-core CPUs. According to the same reason as in section 3.1.3, the type-\uppercase\expandafter{\romannumeral4} computer system $\stackrel{UPM}\Longrightarrow$ type-\uppercase\expandafter{\romannumeral3} computer system. By section 3.1.5, the type-\uppercase\expandafter{\romannumeral3} computer system $\stackrel{UPM}\Longrightarrow$ a type-\uppercase\expandafter{\romannumeral1} computer system. Therefore, the type-\uppercase\expandafter{\romannumeral4} computer system $\stackrel{UPM}\Longrightarrow$ the type-\uppercase\expandafter{\romannumeral1} computer system.

\subsubsection{Type-$\rm\uppercase\expandafter{\romannumeral4}^+$ Computer Systems $\stackrel{UPM}\Longrightarrow$ Type-\uppercase\expandafter{\romannumeral1} Computer Systems}

A type-$\rm\uppercase\expandafter{\romannumeral4}^+$ computer system is composed of a type-\uppercase\expandafter{\romannumeral4} computer system and accelerators. According to the same reason as in section 3.1.2, The type-$\rm\uppercase\expandafter{\romannumeral4}^+$ computer system  $\stackrel{UPM}\Longrightarrow$ a type-\uppercase\expandafter{\romannumeral4} computer system. By  section 3.1.7, the type-\uppercase\expandafter{\romannumeral4} computer system $\stackrel{UPM}\Longrightarrow$ a type-\uppercase\expandafter{\romannumeral1} computer system. Therefore, the type-$\rm\uppercase\expandafter{\romannumeral4}^+$ computer system $\stackrel{UPM}\Longrightarrow$ the type-\uppercase\expandafter{\romannumeral1} computer system.

\subsection{Programming of Various Computer Systems Is Built on One API by UPM}

The above analysis shows that every type of computer system, from type-$\rm\uppercase\expandafter{\romannumeral1}^+$ computer systems to type-$\rm\uppercase\expandafter{\romannumeral4}^+$ computer systems, can be reduced to a type-\uppercase\expandafter{\romannumeral1} computer system, a simplest computer system containing only a single-core CPU, by UMP, and so all types of computer systems are unified into one by UPM in this sense, and so UPM can truly build the programming of various computer systems on one API (i.e. extended file managements of common programming languages for type-\uppercase\expandafter{\romannumeral1} computer systems). The programming of the type-\uppercase\expandafter{\romannumeral1} computer system is the simplest, therefore, UPM can make the programming for various computer systems easier.

\section{Coupled Applications Computing by UPM}

Today, coupled applications computing (like multidisciplinary simulations) plays a key role in many fields. However, It usually requires additional software to provide support to couple multiple applications [3].  This makes coupled appications computing complicated. Here, we present a more natual and simpler approach to coupled applications computing, which is based on UPM. \\

\noindent Assume that a coupled applications computing consists of $n$ applications $A_1$, $A_2$, ..., $A_n$, which are based on MPI, and runs on $n$ clusters cluster$_1$, cluster$_2$, ..., cluster$_n$, respectively. See  Figur 4.

\begin{figure}[htbp]
\centering
\includegraphics[scale=.38]{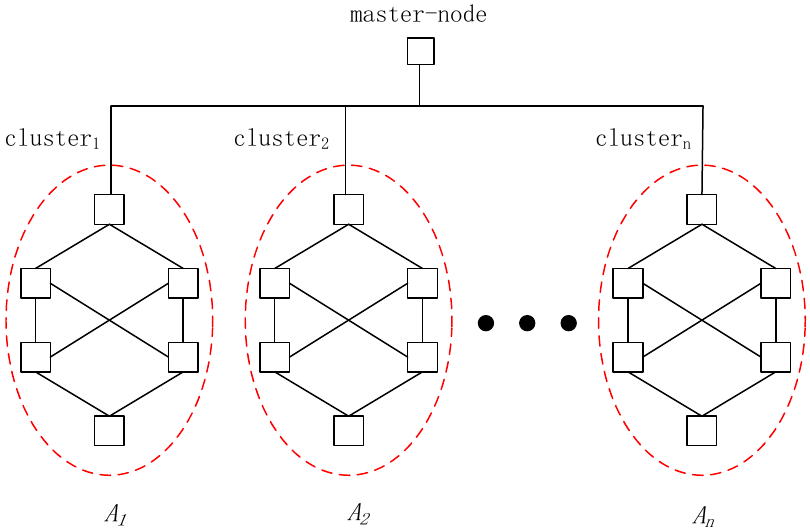}
\caption{A coupled applications computing.}
\end{figure}

\noindent According to section 2.3, in UPM, cluster$_i$ (where $1\le i \le n$) can be imagined to be composed of a node (one CPU) for programs of common programming languages and one “cluster-of-CPU accelerator” for programs based on MPI. We can regard the “cluster-of-CPU accelerator (cluster$_i$ +  $A_i$)” as a general printer (general-printer$_i$), and extend file managements in common programming languages and MPI-I/O to include it as one kind of I/O devices (i.e. files), see Figure 5.\\

\begin{figure}[htbp]
\centering
\includegraphics[scale=.4]{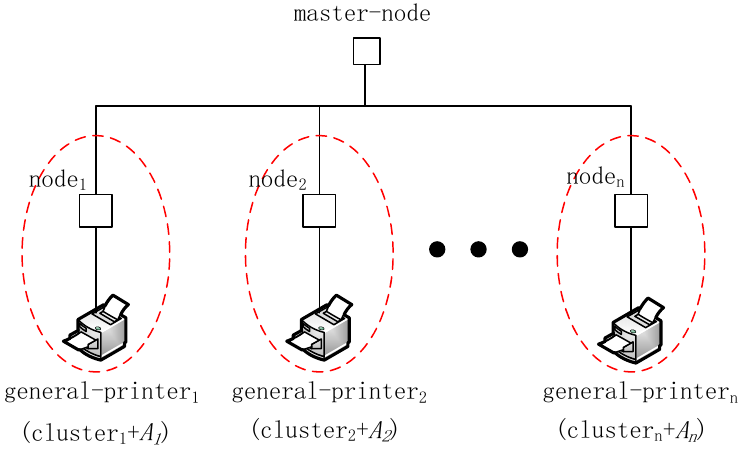}
\caption{The clusters as general printers in the coupled applications computing by UPM.}
\end{figure}

\noindent Now we form a new cluster consisting of master-node, node$_1$, node$_2$, ..., node$_n$, as shown in the blue circle in Figure 6. We call the cluster cluster$_{super}$. Assume that  $A_{super}$, a program based on MPI,  runs on the  cluster$_{super}$.  The $A_{super}$ is responsible for coordinating the $n$ applications $A_1$, $A_2$, ..., $A_n$.\\

\begin{figure}[htbp]
\centering
\includegraphics[scale=.64]{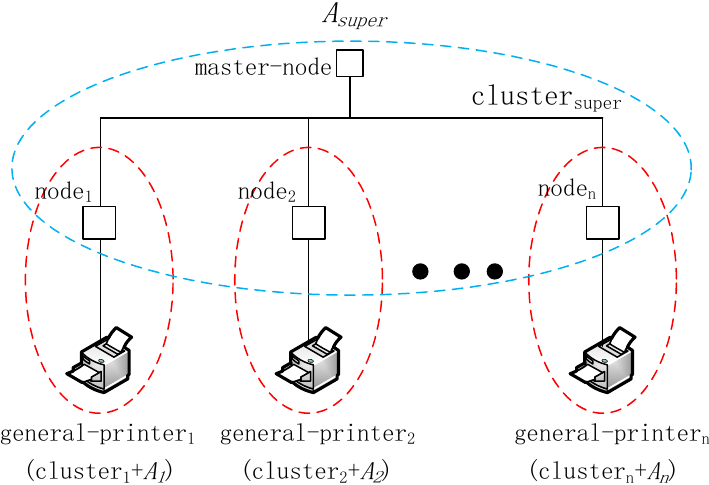}
\caption{cluster$_{super}$ consisting of  master-node, node$_1$, node$_2$, ..., node$_n$.}
\end{figure}

\noindent When $A_i$ sends data to $A_j$ (where $1\le i,j\le n$), the communication is carried out as follows:
\begin{description}
\item[Step 1] $A_{super}$ gets the data in node$_i$ through MPI-I/O from general-printer$_i$ ($A_i$ exactly);
\item[Step 2] $A_{super}$ sends the data from node$_i$ to node$_j$ through MPI over the cluster$_{super}$;
\item[Step 3] $A_{super}$ sends the data in node$_j$ to general-printer$_j$ ($A_j$ exactly) through MPI-I/O. 
\end{description}


\noindent Usually  $A_{super}$ can not communicate with $A_i$ (where $1\le i\le n$) since they are in diffferent parallel MPI worlds. But in UPM,  $A_i$ is a driver of a I/O device (a general printer), and its MPI world is subordinate to (through MPI-I/O), rather than parallel to, the MPI world of $A_{super}$.\\

\noindent The UPM makes coupled applications computing more natural and easier since it only relies on its own power to couple multiple applications through MPI.

\section{Scope of Application of UPM}
Architectures composed of CPUs and accelerators are increasingly popular in computing systems. Today many computing systems, ranging from embedded systems, to mobile computing systems, to HPC systems,  to cloud computing systems, and to quantum-classical hybrid computing systems are equipped with accelerators. UPM can be applied to all these systems since CPU and accelerator technologies are used in the computing systems.

\section{Conclusions and Future Work}

\noindent The key idea of UPM is that an accelerator and a program executting on it can be regarded as a general printer. We can get the following three results from this:

\begin{itemize}
\item This makes any comper system a single-core CPU computer system, which makes programming of any computer systems simpler.
\item This allows OpenMP/Pthread-based code to be separated from the entire program code, resulting in higher modularity. 
\item This can provide another new approach to coupled applications computing.
\end{itemize}

\noindent The above means that UPM is similar to an axiom system, so it is possible that UPM may provide more new solutions to other problems.

\section*{Acknowledgment}

The author would like to thank Dr. Pavan Balaji at Facebook AI. The author was invited by him to visit Mathematics and Computer Science Division at Argonne National Laboratory in 2013, and the idea that GPUs are regarded as one kind of I/O devices was formed preliminarily at that time.



\end{document}